\documentclass[doublecol]{epl2} 
\newcommand{\beq}{\begin{equation}}
\newcommand{\eeq}{\end{equation}}
\newcommand{\beqa}{\begin{eqnarray}}
\newcommand{\eeqa}{\end{eqnarray}}

\usepackage{xspace}
\newcommand{\ecoli}{{\em E.~coli}\xspace}
\usepackage{graphicx}
\usepackage{euscript}
\usepackage{oldgerm}
\usepackage{subfigure}
\usepackage{color}
\usepackage{amssymb}
\usepackage{amssymb,amsfonts,amsmath}

\graphicspath{{./}{../figs_new/}}
\begin{document}

\title{Run-and-Tumble Dynamics of Self-Propelled Particles in Confinement}
\date{\today}

\author{Jens Elgeti and Gerhard Gompper}
\institute{                    
%\affiliation{
Theoretical Soft Matter and Biophysics,
Institute of Complex Systems and Institute for Advanced Simulation,
Forschungszentrum J\"ulich, 52425 J\"ulich, Germany 
}

\abstract{
Run-and-tumble dynamics is a wide-spread mechanism of
swimming bacteria.
The accumulation of run-and-tumble microswimmers near 
impermeable surfaces is studied theoretically and numerically 
in the low-density limit in two and three spatial dimensions. 
Both uni-modal and exponential distributions of the run lengths
are considered. Constant run lengths lead to
{peaks and depletions regions} in the density distribution of
particles near the surface, in contrast to {exponentially-distributed
run lengths}.
Finally, we present a universal accumulation law for large channel widths,
which applies not only to run-and-tumble swimmers, but also to many 
other kinds of self-propelled particles.
}
%\pacs{82.70.-y,45.50.-j,05.40.-a}


\maketitle

\section{Introduction}
Swimming bacteria like \ecoli and Salmonella, with a body length of 
just a few micrometers, are too small for spatial sensing of a stimulus
gradient along their body size \cite{berg77,duse97}. Therefore, they have to 
resort to temporal sensing, where the gradient is determined along the swimming
trajectory. These bacteria have developed a procedure of intriguing
simplicity for chemotactic motion --- 
they perform a run-and-tumble motion, in which nearly straight swimming
segments are interrupted by tumbling events, where the run
length then depends on the sign of the stimulus gradient \cite{berg77,berg04}.
There is an interesting connection of this run-and-tumble dynamics to
L\'evy flights \cite{Thiel2012,Angelani2013}, which suggests 
that this process could be a very efficient search strategy \cite{lomh08}. 

For a dilute suspension of microswimmers in a bulk fluid, 
run-and-tumble dynamics is strictly equivalent to passive-particle 
diffusion for long times.
Here, the diffusion coefficient is given by $D_{eff} = v^2/\tau_r$,
where $v$ is the swimming velocity, and $\tau_r$ is the run 
time \cite{Tailleur2009}. The effective diffusion coefficient is 
typically
much larger than the thermal diffusion coefficient. The equivalence also
holds in the presence of a slowly varying external potential. 
This implies that active Brownian particles (ABPs), which
display a rotational diffusion instead of tumbling events, are
equivalent to run-and-tumble particles (RTPs) under these conditions. 

In fact, the equivalence of ABPs and RTPs has been discovered much 
earlier for the mathematically equivalent case of the conformations of 
semi-flexible polymers. Here, the 
worm-like chain model corresponds to ABPs, whereas the freely-jointed
chain model corresponds to RTPs. The equivalence of the two is
expressed by the Kuhn length $\xi_K$, which is the segment length of
the freely-jointed chain, to equal twice the persistence length 
$\xi_p$ of the worm-like chain, such that the end-to-end distance is 
the same in both descriptions \cite{rubi03}. 

At higher densities of microswimmers, a density-dependent motility 
can cause phase separation and accumulation of both RTPs 
\cite{Cates2013,Paoluzzi2013,soto14} and ABPs \cite{fily12b,wyso14,sten14}, 
which indicates that the dynamics of active particles 
is no longer equivalent to passive diffusion. 
Asymmetric potentials can cause rectification of bacterial motion
\cite{Tailleur2009,koum13,Berdakin2013}.
Also, walls and obstacles break the diffusion equivalence, because
microswimmers accumulate at walls, in contrast to passive particles.
Explanations of this surface trapping usually invoke
hydrodynamics \cite{berk08,Spagnolie2012}. Whether it is the detailed 
hydrodynamics of the corkscrew motion of \ecoli flagella \cite{laug06} 
or the snake-like motion of the sperm tail \cite{elge10}, or the 
far-field hydrodynamics of a hydrodynamic dipole \cite{berk08}, 
hydrodynamics {provides} an 
effective attraction toward boundaries \cite{Spagnolie2012,laug09,elge15}. 
However, for \ecoli, noise also plays an important role and may even 
dominate over the rather weak hydrodynamic interactions \cite{dres:11}. 
Furthermore, it has been shown that persistent motion drives swimmers 
to the wall, even in presence of strong orientational
fluctuations {\cite{elge09,li09,elge13b,fily14,cost14,yang14}}. 
For harmonic confinement, accumulation away from the center
has also been found for run-and-tumble particles \cite{Tailleur2009}.
Thus, it is not obvious under which conditions the equivalence between
RTPs, ATPs, and passively diffusing particles holds. We want to 
clarify this question from the point of view of wall accumulation 
and confinement.

In this letter, we investigate the effect of confinement for particles with
a pure run-and-tumble dynamics in the low-density limit and in the 
absence of hydrodynamic interactions, both analytically and numerically. 
The structure of the density patterns 
of RTPs at hard walls is found to depend strongly on the dimensionality of the
accessible space --- between two planar walls in three dimensions (3D), or 
along a surface with lateral confinement in two dimensions (2D) --- and on 
the run-length distribution, both quantitatively and qualitatively. 
Here, the relevant parameter is the
dimensionless ratio between channel width and (average) run length, whereas
propulsion velocity and tumbling frequency only enter indirectly via
the run length. RTPs are predicted to behave quite differently from 
ABPs.
For narrow channels and constant run lengths, the distribution 
of tumbling events develops pronounced
extrema, with a depletion layer near the wall and a maximum at larger
distances determined by the run length.  These structures disappear for
exponentially distributed run lengths. Thus, the behavior depends 
sensitively on the run-length distribution. 
In contrast, for wide channels, we predict a (nearly) universal 
wall-accumulation law for self-propelled particles. 
This wall-accumulation law only relies on symmetries and
dimensional arguments, and thus holds for many different types
of microscopic swimmers.

Wild-type \ecoli have an average run length of $12 \mu$m 
\cite{BERG1972}, which is the same order of magnitude as the channel width
of microfluidic devices used to manipulate and study these bacteria 
\cite{dilu05,hulm08,Berdakin2013a}. Therefore, our results are relevant,
{\em inter alia}, for the design of microfluidic devices for rectification and 
sorting of run-and-tumble bacteria.

\begin{figure}
\begin{center}
\raisebox{0.37cm}{\includegraphics[width=.30\textwidth]{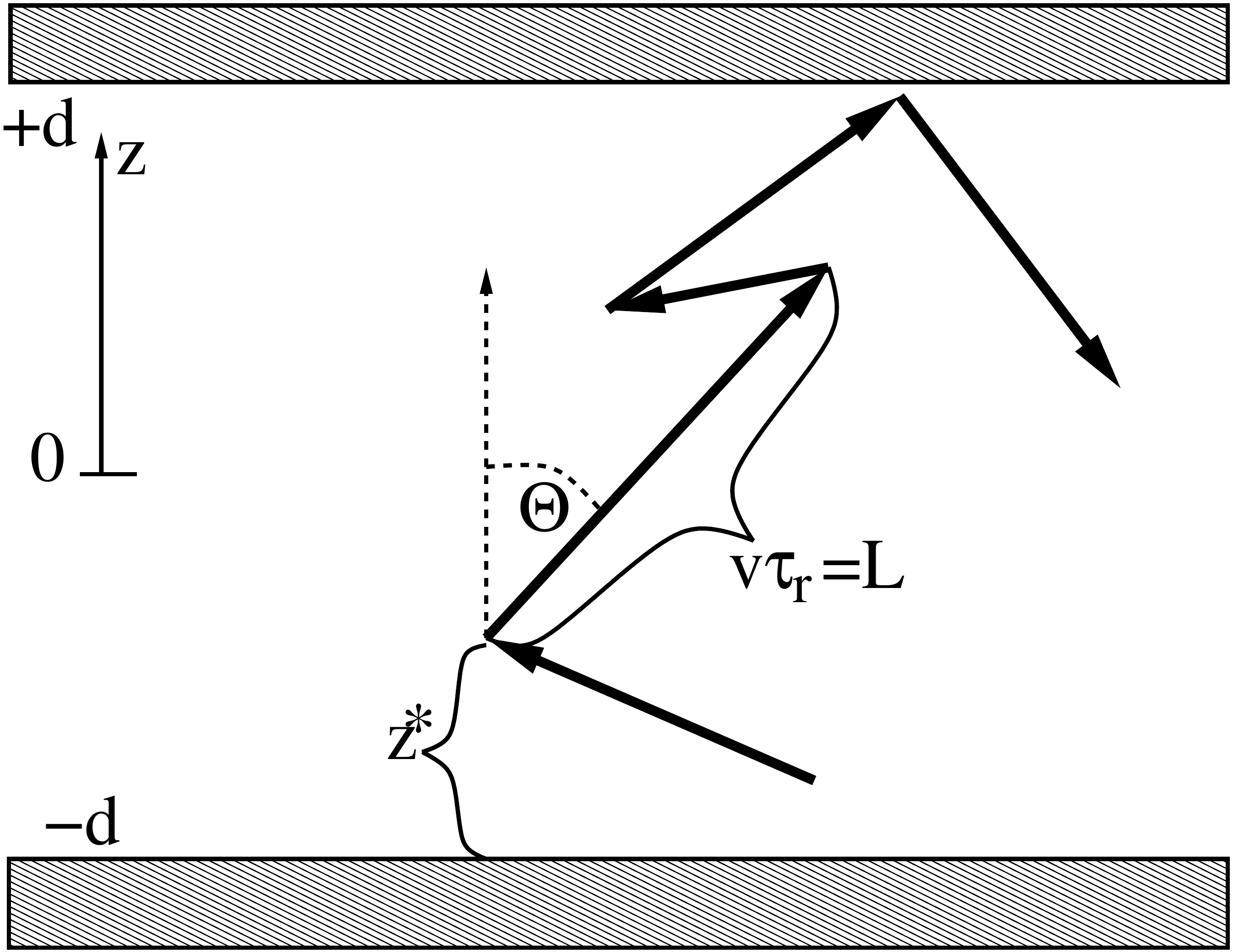}}
\caption{
Schematic of run-and-tumble dynamics.
A ``run'' of the particle with a velocity $v$ for a time
$\tau_r$ is followed by a tumbling event, resulting in a new
orientation $\theta$.  The particle is confined between two parallel 
walls at $z=\pm d$. $z^*$ denotes the distance from the walls.
}
\label{fig:schematic}  
\end{center}
\end{figure}

\section{Model and Simulation Technique}
\begin{figure}
\begin{center}
\includegraphics[width=.40\textwidth]{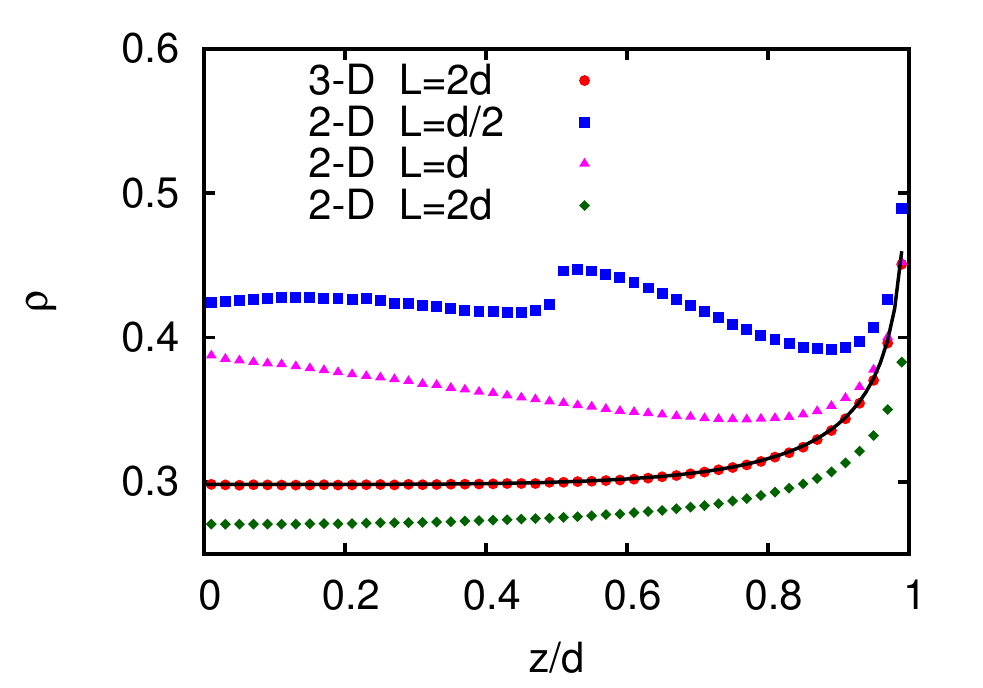}
\caption{
Particle density distribution $\rho(z)$ for various run lengths L. 
The solid line is the exact analytic solution (\ref{eq:3dpartdens}) 
for the particle density in three dimensions for run lengths
larger than the channel width $2d$.
}
\label{fig:density}  
\end{center}
\end{figure}

We study run-and-tumble dynamics of individual microswimmers in
confinement. A particle performs a forward run with a velocity $v$
for a time $\tau_r$. Each run is followed by a tumble event, where a new
orientation angle $\theta$ (see Fig.~\ref{fig:schematic}) is chosen 
randomly on the unit circle (2D) or unit sphere (3D), 
i.e. there is no memory of the orientation before the tumbling
event. The particle 
coordinate $z$ perpendicular to the wall is
then updated by $z(t+\tau_r)=z(t)+\cos(\theta)L$. Here, the run length
$L=v\tau_r$ is either constant, or drawn from an exponential
distribution depending on the dynamics studied. It is important to note
that properties of RTPs do not depend on $v$ and $\tau_r$ separately, but only
on the run length. Due to symmetry, motion parallel to the
wall does not have to be considered. If the particle hits 
a wall, it remains there -- possibly sliding parallel to the wall -- 
until the next tumbling event occurs. After a sufficiently 
long equilibration time, the probability density to find the particle at a
position $z$ is recorded by a histogram over $10^8$ to $10^9$ tumbling steps.
A few examples of density distributions for various run lengths 
are shown in Fig.~\ref{fig:density}.

The trajectory of an RTP is completely defined by the location 
of the tumbling
events, since the motion between these events is just ballistic.
This implies, in particular, that no orientation vector of 
the particle is needed to describe the dynamics. 
Thus, the continuous-time dynamics of a RTP in three
spatial and two orientational dimensions is mapped onto a 
discrete-time-step model in one spatial dimension. 
Physically and mathematically, the fundamental quantity to compute 
in the steady state is then the tumbling density $\phi(z)$. 
The particle density $\rho(z)$ then follows from $\phi(z)$ by a convolution, 
as explained in detail below. Both densities 
are directly asccessible experimentally; however, the tumbling density is 
more difficult to measure, because the trajectories of (all) particles have
to be traced.

Thus, we first focus on the more fundamental tumbling density, which is 
the (normalized) probability to find a tumbling event at a position $z$.
The mirror symmetry of the system is reflected in the symmetry of the 
tumbling density, $\phi(z)=\phi(-z)$. 
The time evolution of the tumbling density is determined by
\begin{align}
  \label{eq:timeevolution}
  \phi(z,t+\tau_r)&=\int_{-d}^{+d} \phi(z',t) p(z-z') dz' .
 \end{align}
Here, $p(\Delta z)$ is the transfer function of
particles moving to a new position, which depends implicitly 
on the run-length distribution. It is the number
of orientational microstates of an unconfined particle which are compatibe 
with a given $\Delta z$-displacement. This probability density
depends on the dimensionality of the system, and on the run length
distribution (unimodal or exponentially distributed).
At the walls, particles accumulate in a $\delta$-distribution 
because all particles that hit the wall are located there.
Thus, we have the boundary conditions
\begin{align}
\label{eq:1}
\phi(\pm d,t) =&0.5 \phi_s(t) \delta(z\pm d)\\
\label{eq:3b} 
0.5 \phi_s(t+\tau_r)=& \int_{-d}^d \phi(z',t) P(-z'-d) dz'   
\end{align}
where $\phi_s$ is the probability to find a tumbling event at the wall,
and $P$ is the cumulative distribution function of $p$,
i.e. $P(z)=\int_{-\infty}^z p(z') dz'$. 

\begin{figure*}
  \centering
  \includegraphics[width=0.315\textwidth]{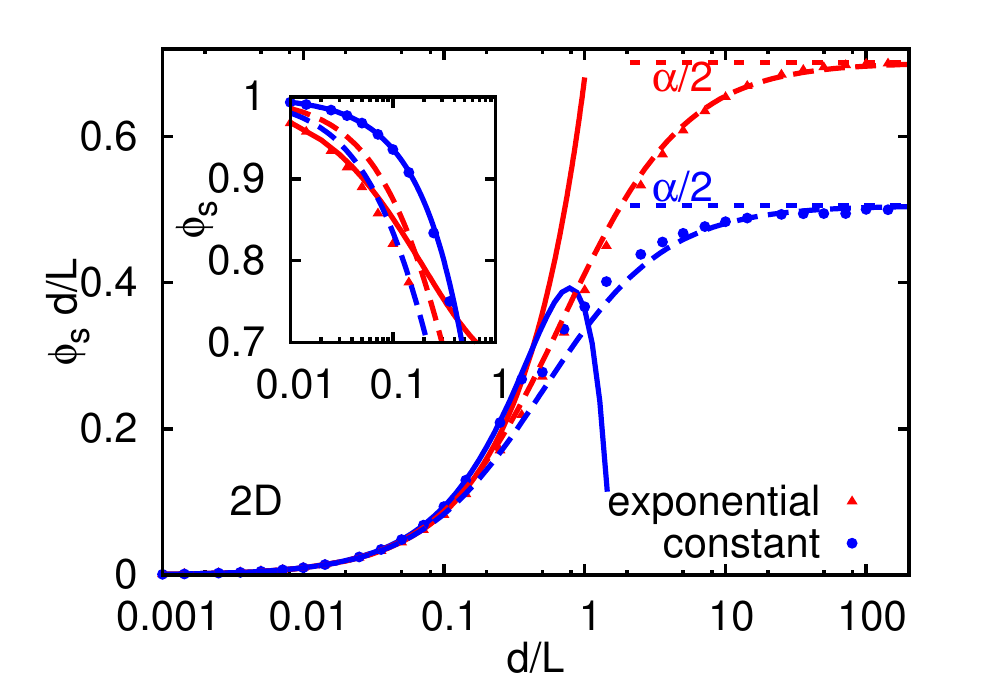}
\includegraphics[width=0.32\textwidth]{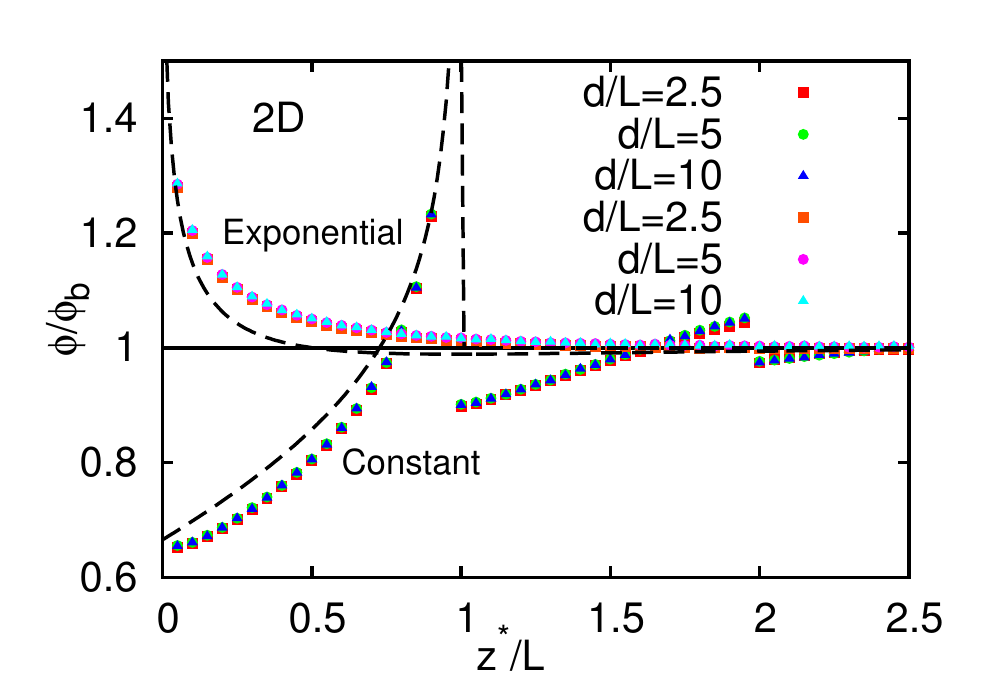}
\includegraphics[width=0.35\textwidth]{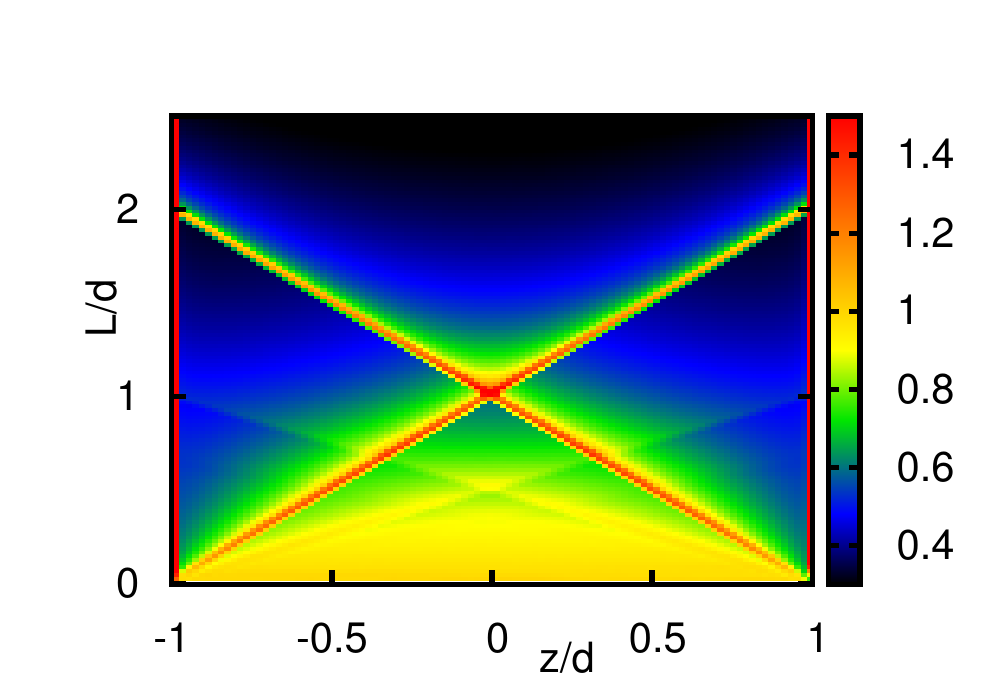} \\
\vspace{-0.7cm}
\includegraphics[width=0.315\textwidth]{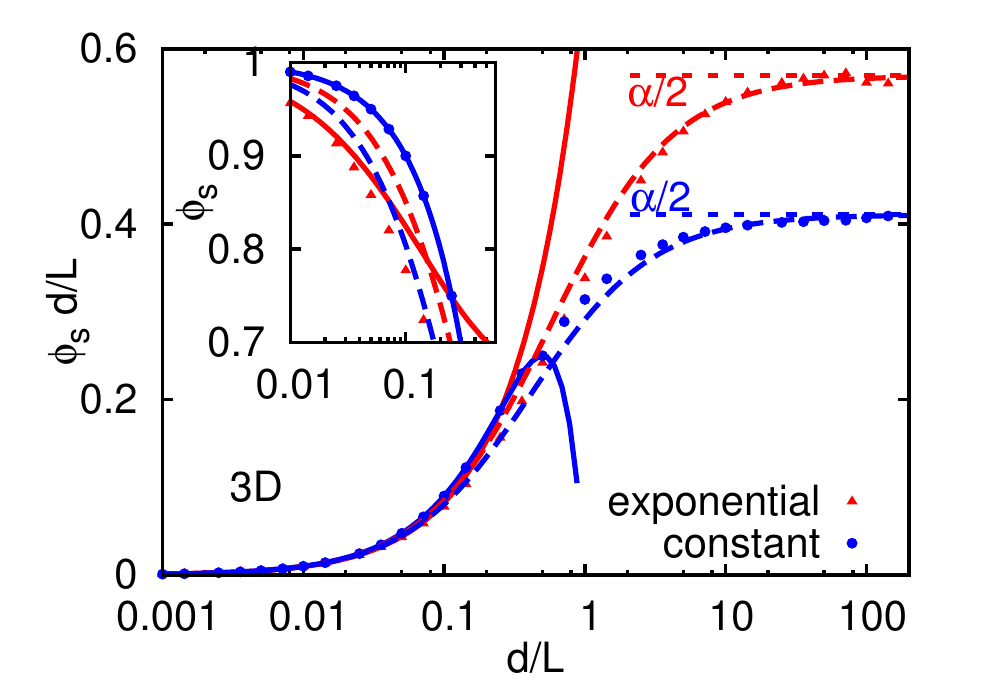}
\includegraphics[width=0.32\textwidth]{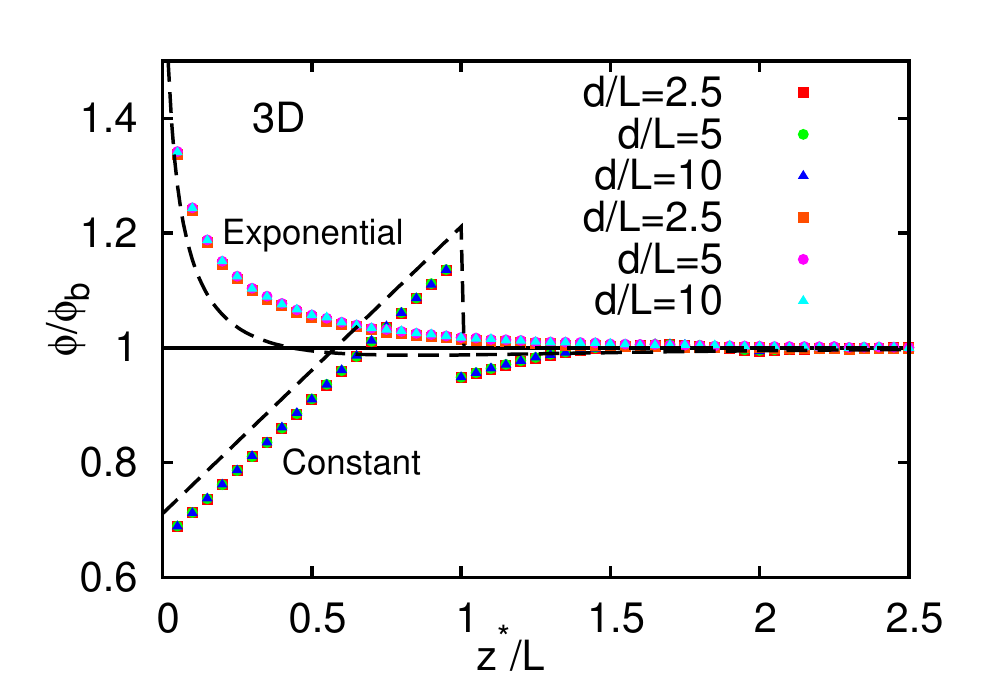}
\includegraphics[width=0.35\textwidth]{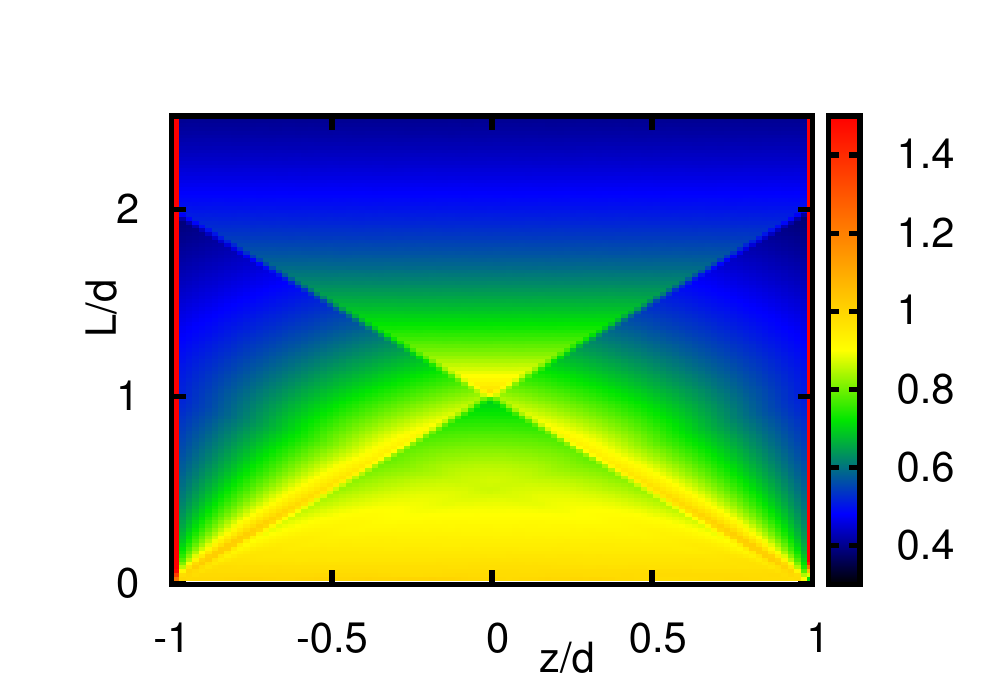}
\caption{Tumbling density profiles $\phi(z)$ for {\bf (top)} two and {\bf (bottom)} 
three dimensions.
{\bf (left)} {Scaled} surface density $\phi_s d/L$ as a function of the 
ratio $d/L$ of channel width and run length. 
Solid lines are analytical approximations for narrow channels (see text), 
dashed lines are fits to the large-channel approximation
(\ref{eq:lchannel}).
{Note that for $d\gg L $, $(\phi_s d/L)$ approaches $\alpha/2$.}
{\bf (center)}
Scaled tumbling density $\phi(z^*)/\phi_b$ as a function of the scaled 
distance  $z^*/L=(z+d)/L$ from the wall. Dashed lines show the approximation 
{$\phi_1(z^*)$, which is obtained from a first-order iteration of 
Eq.~(\ref{eq:timeevolution}) (see text). }
{\bf (right)} 
Density of tumbling events $\phi(z/L)$ inside the channel for
various (constant) run lengths.
For {$2d \ge L$}, the presence of the two walls induces strong density modulations. 
}
\label{fig:phi}
\end{figure*}

\section{Constant Run Lengths} 
We begin our analysis in the simplest case of constant run length $L$. 
The transfer function is discontinuous at the run length, as runs longer than $L$
cannot occur. 
In three dimensions, the transfer function $p(z)dz$ is obtained by an 
integral over the surface of a sphere of radius $L$ with values of the vertical
displacement between $z$ and $z+dz$. This yields immediately that the transfer 
function is 
\begin{align}
\label{eq:9}
  p_{(3,c)}(z)&=\frac{1}{2L} \Theta(L-z) \Theta(L+z), 
\end{align}
where $\Theta(z)$ is the Heaviside step function. Subscripts denote 
dimensionality and the type of run length distribution ($c$=constant, 
$e$=exponential).  Similarly, in two dimensions, integration over 
a circle of radius $L$ with displacement between $z$ and $z+dz$ yields 
even a divergence of the transfer funtion at the run length, 
\begin{align}
  \label{eq:transp}
  p_{(2,c)}(z)=&\frac{1}{\pi L \sqrt{1-(z/L)^2}} \Theta(L-z) \Theta(L+z).
\end{align}
The simplicity of the 3D transfer function allows for
an analytic solution for narrow channels with $2d<L$,
\begin{align}
  \label{eq:3d}
  \phi_{(3,c)}(z)=\frac{1}{2L}+\frac{1-d/L}{2}[\delta(z-d)+\delta(z+d)]
\end{align}
In 2D, an analytical solution can only be obtained 
by assuming that the number of tumbling events in the bulk is
negligible compared to tumbling events at the wall 
(i.e. $\int_{-d+\epsilon}^{d-\epsilon}\phi(z)dz \ll \phi_s$ with an
infinitesimal length $\epsilon$). 
This results in  
\begin{align}
  \label{eq:2d}
\phi_{(2,c)}(z)=&\frac{\pi}{\pi+\sin^{-1}(2d/L)}
      \left[ 0.5\delta(z-d)+0.5\delta(z+d)\right] \nonumber \\
  &+\frac{1}{2\pi+2\pi\sin^{-1}(2d/L)}\times \nonumber\\
     & \left[ \frac{1}{\sqrt{L^2-(z-d)^2}} +\frac{1}{\sqrt{L^2-(z+d)^2}}
      \right] .
\end{align}

These analytical results and corresponding simulation data are displayed
in Fig.~\ref{fig:phi}. The comparison shows that  
the solution (\ref{eq:3d}) in 3D and the approximate expression (\ref{eq:2d}) 
in 2D work very well for the appropriate regimes.
Figure~\ref{fig:phi} reveals
that the walls induce a very rich structure 
of the tumbling density in the channel for $2d \ge L$, i.e. for channels  
wider than the run length. 
The density profiles all collapse onto a single master curve when 
the tumbling density is scaled by the bulk density $\phi_b$ (the 
density far away from the wall) and distances are scaled by the run length.
In this case, the high particle density at the walls 
generates depletion regions near the walls, and two pronounced peaks 
at a distance $L$ from the wall, which can easily be recognized in 
Fig.~\ref{fig:phi} (middle) and (right). In 2D, these primary peaks generate
secondary peaks for $d \ge L$, which are again displaced by a 
distance $L$ further away from wall they first came from. In 3D, the 
primary singularities are too weak
to generate visible secondary peaks. All bulk singularities disappear for
$L>2d$, compare Eq.~(\ref{eq:3d}), because particles can move {directly} 
from one wall to the other in a single step. 
These depletion zones and peaks can be understood by starting from a 
uniform bulk distribution plus $\delta$-peaks at the walls, and iterating 
Eq.~(\ref{eq:timeevolution}) once, as explained in more detail below.

\section{Exponential Run-Length Distribution}
In the case of a distribution of run lengths, the transfer functions 
are obtained by convolution of
the (conditional) probabilities $p(z|L')$ --- resulting from step with 
run length $L'$ --- with the run-length distribution $p_{len}(L')$, i.e.
\begin{align}
p_{(n,len)}(z)=&\int_z^\infty  P_{(n,c)}(z|L') p_{len}(L')dL'
\end{align}
The integral has a lower boundary at $z$, because $\Delta z=z$ cannot be 
achieved with $L'<z$. 

We focus here on exponential run length distributions,
\begin{equation}
p_{len}(L') = \lambda \exp(-\lambda L'), 
\end{equation}
with $\langle L' \rangle \equiv L = 1/\lambda $, which mimic the run-length distribution
of {\em E. coli}.
This yields the transfer functions
\begin{align}
  \label{eq:16}
  p_{(2,e)}(z) =& \frac{1}{\pi L}  K_0\left(\frac{|z|}{L}\right)  \ , \nonumber
  \\ 
  p_{(3,e)}(z) =& \frac{1}{2 L}E_1\left(\frac{|z|}{L}\right)
\end{align}
with $K_0$ a Bessel function and $E_1$ an exponential-integral function.
The continuity and strong decay of these transfer functions leads to the
disappearance of all the singularities of the tumbling density found for constant 
run length, see Fig.~\ref{fig:phi}. 
{Thus, the tumbling density is highly sensitive to the run-length
distribution.}

We estimate the
density profile in small channels, by neglecting the particles in the
bulk in Eq.~(\ref{eq:timeevolution}), as for constant run length.
In two and three spatial dimensions, we then obtain 
\begin{align}
\phi_{s(2,e)} &= \left[1+\lambda d K_0(2\lambda d)L_{-1}(2\lambda d) \right. \nonumber \\
            &  \left. \ \ \ \ \ +\lambda d K_1(2\lambda d)L_{0}(2\lambda d)\right]^{-1}\\
\phi_{s(3,e)} &= \left[1+\frac{1-E_2(2\lambda d)}{2} \right]^{-1}
\end{align}
where the $L_i(x)$ with $i=-1,0$ are Struve functions, and $E_i(x)$ with $i=1,2$ are 
exponential-integral functions.
Figure~\ref{fig:phi} demonstrates that 
these approximations work very well for narrow channels. 
However, the critical length scale, where they break down, is
significantly lower than for constant run length. 
The reason is that 
the transfer functions for exponentially distributed run lengths decay
much faster than the transfer functions for constant run lengths. This implies that 
there are more particles inside the channel, so that the approximation
$\int_{-d+\epsilon}^{d-\epsilon}\phi(z)dz \ll \phi_s$ breaks down
already at smaller channel widths.

\section{Scaling Behavior of Wall Density} 
For channels much wider than the (average) run length, we can use 
scaling arguments to determine the wall accumulation of particles.
For $d \gg L$, the tumbling density profile eventually becomes flat far from the walls.
Everything else fixed, the surface density has to be linear in the
bulk density $\phi_b$ defined as the (constant) density far from the
walls.
Since the (average) run length $L$ is the only {relevant} length
scale near the wall, the proportionality factor has to be linear in $L$, so that 
\footnote{Alternatively, it can be argued that the surface density $\phi_s$,
          which is dimensionless in our description, can only depend on the
          ratio of the two available length scales $L$ and $d$, which implies
          $\phi_s = F(d/L)$, with some unknown scaling function $F(x)$.

          This can also be seen by considering an infinite half-space, with
          a wall at $z=0$. In this case, the boundary condition is that the density
          approaches $\phi_b$ for $z \to \infty$. Then, $L$ is the only available
          length scale.  For finite but very wide channel, the density profile should
          not change. However, the normalization of the probability density introduces
          a constraint on $\phi_b$, which implies $\phi_b\sim 1/d$.}
\begin{align}
   \phi_s = \alpha L \phi_b
\end{align}
with a dimensionless prefactor $\alpha$. In a channel of finite
with $d$, normalization then gives 
\begin{align}
  \label{eq:lchannel}
    \phi_s=\frac{\alpha L}{\alpha L + 2d}= \frac{1}{1 + 2d/(\alpha L)} .
\end{align}
The surface accumulation factor $\alpha$ is
independent of run length, and only depends on the dynamics (i.e. 2D/3D,
constant run length/ exponential run-length distribution).
From our simulations, we obtain the accumulation 
factors shown in Table \ref{tab1}.
The large-distance approximation (\ref{eq:lchannel}) works very well 
for $d>L$, and even 
for smaller channels it is not too far off (see Fig.~\ref{fig:phi}).
Unimodal and exponential run-length distributions result in 
accumulation factors $\alpha$, which are clearly different, 
but still of the same order of magnitude. Thus, measurements of 
$\alpha\equiv \phi_s/(L\phi_b)$ for $d \gg L$
might provide a new possibility to characterize run-length distributions
experimentally. 

\begin{table}
\begin{tabular}{|c|c|c|}
  \hline
$\alpha$ & 2D & 3D \\
\hline
RTPs, constant $L$                  &\ 1.01(2) \  & \ 0.82(2) \ \\
RTPs, exponential $L'$ distr.        &\ 1.40(2) \  & \ 1.14(2) \ \\
\hline
ABPs, $L=2\xi_p$    &  \ 0.80(3) \   & \ 0.37(3) \ \\
\hline
\end{tabular}
\caption{Accumulation factor $\alpha$ for various self-propelled particles
in two and three spatial dimensions. For ABPs, $\alpha$ values are obtained 
from direct Langevin simulations; in this case, we employ the ``Kuhn" length
as the characteristic length scale.}
\label{tab1}
\end{table}

\begin{figure}
\centering
\includegraphics[width=0.40\textwidth]{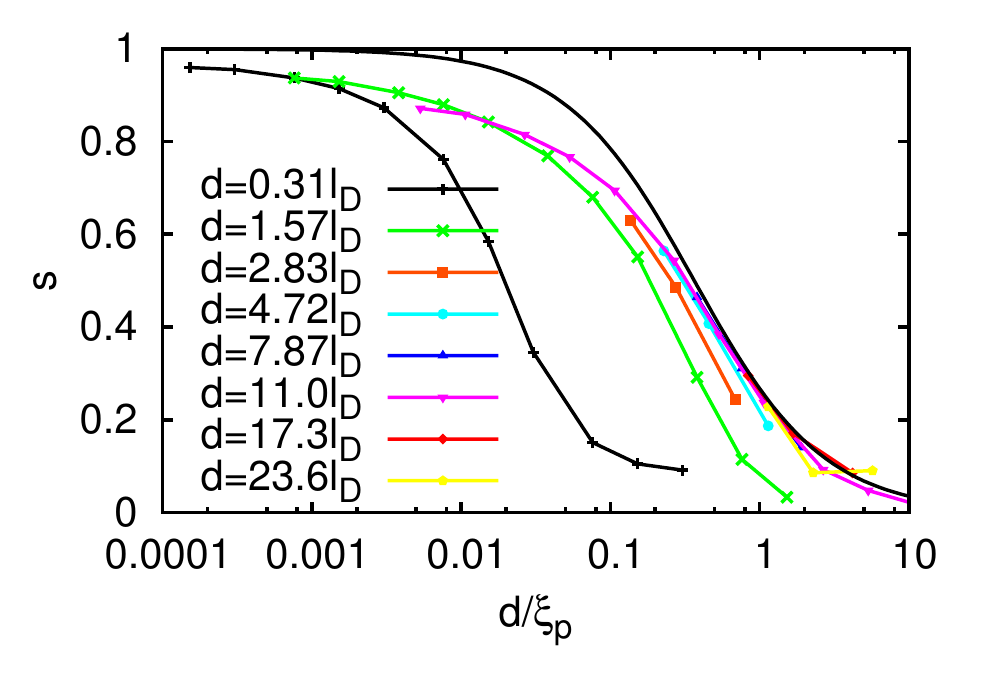}
\caption{Surface excess $s=\int(\rho(z)-\rho_b)dz$ for 
self-propelled Brownian spheres. Data from Ref.~\cite{elge13b}. 
Here, $s$ is to be considered equivalent to $\phi_s$.
The wide-channel approximation (\ref{eq:lchannel}) works already 
rather well for channels widths $2d$ comparable to the persistence 
length $\xi_p$, as long as $\xi_p$ is much larger than the diffusive 
length scale $l_D=\sqrt{D_t/D_r}$.
}
\label{fig:rhos}
\end{figure}

Since these arguments rely only on dimensional analysis, the results should be 
valid for other types of 
self-propelled particles as well (as long as there is one dominant 
length scale of the dynamics). 
For active Brownian spheres, this length scale is the
persistence length of the trajectory $\xi_p=v/D_r$, 
where $D_r$ is the rotational diffusion coefficient.
Using data from Ref.~\cite{elge13b}, we find indeed an excellent
agreement for channels much larger than the diffusive length scale
$l_D=\sqrt{D_t/D_r}$, where $D_t$ is the translational diffusion
coefficient (see Fig.~\ref{fig:rhos}).
{Note that Eq.~(\ref{eq:lchannel}) also predicts a crossover 
from narrow- to wide-channel behavior at $L \simeq 2d$ for all kinds of 
self-propelled particles.} 
The fact that the $\alpha$-values in Table \ref{tab1} differ 
significantly for RTPs and ABPs clearly demonstrates that these two 
types of self-propelled motion are {\em not} equivalent near surfaces.
However, the fact that these factors are all of order unity 
emphasizes the generic aspect of wall accumulation.

\section{Near-Wall Density in Wide Channels} 
For wider channels, the surface density $\phi_s$ is well described by
Eq.~\ref{eq:lchannel}.
To understand the structure of the density distribution close to the wall,
the stationary form of Eqs.~(1) to (3) 
can be used to obtain an analytical approximation.
We start as an initial guess with a $\delta$-distribution at the wall, 
with an amplitude $\phi_s$, plus a constant tumbling density $\phi_b$ in 
the bulk (see Eq.~\ref{eq:lchannel}).  An iteration with Eq.~(1) then yields
\begin{align}
  \phi_1(z^*) &= \frac{\phi_s}{4}\delta(z^*) 
      + \phi_b\delta(z^*)\int_0^\infty P(z') dz' \nonumber \\
     + \frac{\phi_s}{2}&p(z^*)
      + \phi_b\left(1-\int_\infty^0 p(z^*-z')\text{d}z' \right).
\label{eq:dip}
\end{align}
Note that the last two terms on the right-hand side of Eq.~(\ref{eq:dip}) 
determine the spatial dependence of the tumbling
density near the wall. 
Figure~\ref{fig:phi} (center) shows that 
this first-order calculation can qualitatively explain 
the numerical results for the structure of the tumbling-density profile. 
In particular, for constant run lengths, 
it reproduces and explains the near-wall dip in the tumbling density, 
i.e. the formation of a depletion layer close to the wall, and in turn
a peak and discontinuity at $z=L$. 
As shown in Fig.~\ref{fig:phi} (right), this peak
leads to interesting patterns in the tumbling-density distribution for
channel widths larger than the run length, in particular a very pronounced
peak in the channel center for $d=L$.

\section{Microswimmer Density} 
Finally, we connect the tumbling density to the number density of microswimmers. 
This requires the convolution of
the tumbling density with the spreading function $f(\Delta z)$ 
of one run, 
\begin{align}
  \label{eq:3}
  \rho(z)&=\int_{-\infty}^\infty \phi(z')f(z-z') \text{d}z'
\end{align}
(where the particles which would penetrate the walls have to be 
``folded back" to the wall, i.e. $\phi_s(d)=\int_d^\infty P(z) dz$).
For constant run length, the spreading function $f(\Delta z)$ is 
obtained from the transfer functions as 
\begin{align}
  \label{eq:10} 
f(z)&=\int_z^\infty \frac{p(z')}{z'}\text{d}z'
         \ \ \ {\rm for} \ \ z>0,
\end{align}
and $z<0$ follows from symmetry
\footnote{In the more general case, Eq.~(\ref{eq:10}) has to me
modified to account for the run-length distribution }.
Note that if the tumbling time $\tau_{t}$ cannot be neglected compared 
to the run time $\tau_r$, the tumbling density has to be added 
proportional to $\tau_{t}/\tau_{r}$ in Eq.~(\ref{eq:3}).
As an example, we consider here the case of thin three-dimensional 
channels ($2d<L$) and constant run length, which can be solved 
analytically. Here, the spreading function is found to be
\begin{align}
  f(z)=\frac{1}{2L}\ln\left(\frac{L}{|z|}\right), 
\end{align}
which yields the particle density (for $z>0$)
\begin{align}
\label{eq:3dpartdens}
\rho(z)=&\frac{1}{4L}
   \left[\ln\left(\frac{L}{d+z}\right)+\ln\left(\frac{L}{d-z}\right)\right]
                                                 \nonumber \\
  &+\frac{d}{4L^2}
       \left[2 +\frac{z}{d}\ln\left(\frac{d-z}{d+z}\right) \right] \nonumber \\
  &+\frac{\rho_s}{2}\delta(z-d) +\frac{\rho_s}{2}\delta(z+d)
\end{align}
where the surface density $\rho_s$ of particles is obtained from  
the normalization condition.
Equation (\ref{eq:3dpartdens}) fits the simulations perfectly, 
without any adjustable parameters, see Fig.~\ref{fig:density}.

\section{Conclusions}
{We have shown that run-and-tumble dynamics of self-propelled particles 
leads to highly structured density distributions near impenetrable surfaces.} 
Due to the absence of 
translational diffusion, accumulation
materializes in the form of $\delta$-function peaks at the surface. 
Diffusion would broaden these peaks, similarly as predicted for 
ABPs \cite{elge13b}. Close to confining walls, RTPs are thus clearly not equivalent 
to either diffusing particles or ABPs. 
The density distributions are predicted to depend sensitively on the spatial
dimensionality and on the run-length distribution, where the typical
length scale is set by the (average) run length.

While the dynamics considered here is certainly oversimplified for
real microswimmers like 
\ecoli, it captures the essential aspects of run-and-tumble motion, 
and similar results can be expected for 
other types of Levy flights. In particular, the limit of large wall
separations for the accumulation is very generic, and should thus apply to many
systems of self-propelled particles and microswimmers \cite{laug09,elge15}. 
It will be interesting to see whether this behavior extends to systems
in which hydrodynamic interactions play a significant role.

Another interesting issue is the behavior of RTPs at finite particle
density \cite{Cates2013,Paoluzzi2013,soto14}. For high density, the characteristic
features in confinement revealed by our study will almost certainly be washed out,
because collisions will dominate over tumbling events. However, interesting
behavior can be expected in confinement, when the average distance between
particles becomes comparable with the run length.  

%\bibliographystyle{eplbib} 
%\bibliography{micro.bib}

\end{document}